\def\year{2022}\relax
\documentclass[letterpaper]{article} 
\usepackage{aaai22}  
\usepackage{times}  
\usepackage{helvet}  

\usepackage{courier}  
\usepackage[hyphens]{url}  
\usepackage{graphicx} 
\usepackage{natbib}  
\usepackage{caption} 
\usepackage[subrefformat=parens]{subcaption}

\DeclareCaptionStyle{ruled}{labelfont=normalfont,labelsep=colon,strut=off} 
\usepackage{color}
\usepackage{algorithm}
\usepackage{amsmath,amssymb,amsfonts}
\usepackage{algpseudocode}

\usepackage{newfloat}
\usepackage{listings}
\floatstyle{ruled}
\newfloat{listing}{tb}{lst}{}
\floatname{listing}{Listing}

\title{First to Possess His Statistics: Data-Free Model Extraction Attack on Tabular Data}
\author{
    Masataka Tasumi,\textsuperscript{\rm 1}
    Kazuki Iwahana,\textsuperscript{\rm 1}
    Naoto Yanai,\textsuperscript{\rm 1}
    Katsunari Shishido,\textsuperscript{\rm 2}
    Toshiya Shimizu,\textsuperscript{\rm 2}\\
    Yuji Higuchi,\textsuperscript{\rm 2}
    Ikuya Morikawa\equalcontrib,\textsuperscript{\rm 2}
    Jun Yajima\equalcontrib\textsuperscript{\rm 2}
}

\affiliations{
    \textsuperscript{\rm 1} Osaka University\\

    \textsuperscript{\rm 2} Fujitsu Limited 
}

\usepackage{bibentry}

\begin{document}







\maketitle

\begin{abstract}
Model extraction attacks are a kind of attacks where an adversary obtains a machine learning model whose performance is comparable with one of the victim model through queries and their results. This paper presents a novel model extraction attack, named TEMPEST, applicable on tabular data under a practical data-free setting. 
Whereas model extraction is more challenging on tabular data due to normalization, TEMPEST no longer needs initial samples that previous attacks require; instead, it makes use of publicly available statistics to generate query samples. Experiments show that our attack can achieve the same level of performance as the previous attacks. 
Moreover, we identify that the use of mean and variance as statistics for query generation and the use of the same normalization process as the victim model can improve the performance of our attack. 
We also discuss a possibility whereby TEMPEST is executed in the real world through an experiment with a medical diagnosis dataset. 
We plan to release the source code for reproducibility and a reference to subsequent works. 
 \end{abstract}

\section{Introduction}  
\label{Introduction}

\subsection{Background}
Machine learning models are becoming more valuable because machine learning models are proven useful in various applications. In general, construction of a useful machine learning model is very costly, mainly because collecting a large amount of data for training and the training process itself require much effort, computation, and financial cost. For example, construction of a XLNet model~\cite{yang2019xlnet}  costs \$61,000 -- \$250,000\footnote{\url{https://syncedreview.com/2019/06/27/}\\\url{the-staggering-cost-of-training-sota-ai-models/}}.  Hence, many research works in recent years have focused on model extraction attacks, in which an adversary can obtain a duplicate (called substitute model) of a victim machine learning model through query access to it~\cite{PRADA,orekondy2019knockoff,truong2021datafree,pal2020activethief,kariyappa2020maze}. Model extraction attack enables an adversary to effectively construct a substitute model of a victim model, without incurring costs for data collection and model training. Recent literature presents successful attacks against models for various applications, such as natural language processing models~\cite{Krishna2020Thieves,shirish2020thievesmonolingual} and generative models of images~\cite{szyller2021goodartists}. Such attacks may lead to unauthorized use of an expensive model, and revenues of an owner of the victim model are impaired. Therefore, model extraction attack is a serious problem to machine learning applications. 

Previous model extraction attacks~\cite{FFA16,PRADA,orekondy2019knockoff,truong2021datafree,pal2020activethief,kariyappa2020maze} have been often discussed on benchmarks over simple image datasets such as MNIST and CIFAR-10, but evaluations on other data structures including tabular data are rarely discussed. In general, suitable techniques and performances achieved of machine learning models depend heavily on data structure, and we, therefore, expect data structure should also affect the success and failure of model extraction attacks.

In this paper, we demonstrate to answer this question: \textbf{how can we realize a more realistic and practical model extraction attack against models for tabular data?}
Tabular data are the most common data structure in real-world machine learning applications~\cite{arik2021tabnet}, such as a medical dataset consisting of a patient's body weight and blood pressure, a financial dataset consisting of customer's preference and sales. Thus, we expect that investigating the above question is beneficial to a variety of practical machine learning applications. 
However, general-purpose technique to speed up convergence of model training, which is effective on image and natural language data, is not easily applicable on tabular data~\cite{yoon2020vime}. Also, most previous studies focus on image classification models and do not imply that they are applicable to the classification of tabular data.




\subsection{Contribution}
In this paper, we present a novel model extraction attack against a model in the tabular domain, named \textit{TEMPEST (Tabular Extraction from Model by Public and External STatistics)}.
Supposing that an adversary could only access publicly and externally available statistical information (e.g., mean and variance) about the training dataset of the victim model, our model extraction attack is effective against classification models for tabular data. The above assumption can be regarded as a practical version of the data-free setting proposed in recent literature~\cite{truong2021datafree,kariyappa2020maze}, where an adversary does not have any sample from the training dataset of the victim model.
In this sense, our model extraction attack belongs to a novel attack class.

TEMPEST generates adequate query data for model extraction by leveraging publicly available statistical information that includes values, e.g., mean, variance, minimum, and maximum, close to the training dataset. As discussed in Section of Difficulty of Data-free Model Extraction on Tabular Data, this assumption is more realistic than data-free setting in previous studies~\cite{kariyappa2020maze,truong2021datafree}.

We conduct experiments with tabular datasets, and show that TEMPEST can construct substitute models whose performance is comparable with the state-of-the-art attack in PRADA~\cite{PRADA}, which requires initial samples from the victim's training data, and a typical data-free attack. Moreover, we find desirable conditions for TEMPEST's query data generation from publicly available statistical information: use of mean and variance and use of the same normalization as the victim model. They improve the performance of TEMPEST as shown in Section of Discussion. 
We finally show that TEMPEST potentially realizes data-free model extraction in the real world as long as enough statistical information is available for each inference class. 

\section{Related Work} 

Research on model extraction attacks aims to reduce background knowledge of an adversary. 
The first model extraction attack~\cite{FFA16} requires an adversary to own about 20\% training data of a victim model. 
The subsequent works~\cite{Papernot2017,PRADA} presented attacks whereby an adversary estimates data distribution for query generation through adversarial examples~\cite{szegedy2014intriguing} based on a small amount of initial samples from the training dataset. 
Afterward, attacks based on a substitute dataset~\cite{orekondy2019knockoff,pal2020activethief}, whose data distribution is similar to the training data of a victim model, were presented. 
However, it is often difficult to prepare such a substitute dataset. 
In recent years, data-free model extraction attacks~\cite{truong2021datafree,kariyappa2020maze} were presented, where an adversary obtains a substitute model only by querying to a victim model without background knowledge. 
TEMPEST is a more practical and feasible model extraction attack in the data-free setting as described in Section of Difficulty of Data-free Model Extraction on Tabular Data. 



Since the performance of model extraction attacks depends on architectures of victim models~\cite{hu2020deepsniffer}, 
several works have investigated the model extraction on inference tasks~\cite{reith2019efficiently} and training methods~\cite{chen2021stealingrein,chandrasekaran2020exploring,wang2018stealinghyper}. 
In recent years, attacks to real-world applications such as image processing~\cite{szyller2021goodartists} or natural language processing~\cite{Krishna2020Thieves,shirish2020thievesmonolingual,zanella-beguelin21a2021greybox} were shown. 
Moreover, there are attacks~\cite{zhu2021hermes,hu2020deepsniffer,hua2018reverse} that infer model architectures through access to a physical device such as GPU. 
We believe that more advanced model extraction attacks are available in various applications by combining TEMPEST with the attacks mentioned above. 


\section{Problem Setting} 

In this section, we explain basic definitions of machine learning models and model extraction attacks. 

\subsection{Machine Learning and Model Extraction Attack}

Let $\mathbb{N}$ denote a set of natural numbers, $\mathbb{R}$ a set of real numbers, and $\mathcal{C}$ a (finite) set of labels.
For arbitrary natural numbers $m, n \in \mathbb{N}$, a machine learning model is defined as a function $M: \mathbb{R}^m \rightarrow \mathcal{C}$, which takes an input feature vector $x$ and outputs a probability vector $M(x)$.
More precisely, $\{p_i\}_{i \in [1,m]}= M(x)$ represents a vector of probability $p_i$ such that $x$ is classified as the class $c_i \in \mathbb{C}$ for $i \in [1,m]$. 
With these notations, a machine learning model is used in two processes: a training process, which takes pairs of data and label $(x, c) \in \mathbb{R}^m \times \mathcal{C}$ as an input and then updates parameters $\theta$ of $M$, and an inference process, which takes a vector $x'$ as an input and outputs a probability vector $M(x')$.

The goal of an adversary $\mathcal{A}$ for the model extraction attack is to obtain a new machine learning model $M_{\mathcal{A}}$ (called a substitute model) mimicking the function of a victim's model $M_{\mathcal{V}}$ (called a victim model).
To this end, $\mathcal{A}$ queries to $M_{\mathcal{V}}$ with an unlabeled dataset $X_{\mathcal{A}}$ to obtain an inference result $\{p_i\}_{i \in [1,m]}= M_{\mathcal{V}}(x)$ for $x\in X_A$. 
Then, $\mathcal{A}$ obtains $M_{\mathcal{A}}$ with almost the same performance as $M_{\mathcal{V}}$ through training $M_{\mathcal{A}}$ using $X_{\mathcal{A}}$ and $M_{\mathcal{V}}(x)$ for any $x\in X_A$. 

There are two well-known metrics for evaluating the performance of model extraction attacks. For a given test dataset $D_{val}: X \times Y$, a substitute model $M_{\mathcal{A}}$ is evaluated with the following metrics: 

\textbf{Accuracy} is a metric to measure how a substitute model $M_{\mathcal{A}}$ can infer accurately. It is defined as $ \mbox{Pr}_{x_{val} \sim D_{val}} \left[ \mbox{argmax}(M_{\mathcal{A}}(x_{val})) = y \right]$. 
$\mathcal{A}$ can evaluate the performance of $M_{\mathcal{A}}$ by its accuracy.

\textbf{Fidelity} is a metric to measure how outputs of a substitute model $M_{\mathcal{A}}$ and a victim model $M_{\mathcal{V}}$ are identical. It is defined as $ \mbox{Pr}_{x_{val} \sim D_{val}} \left[ \mbox{argmax}(M_\mathcal{A}(x_{val})) = \mbox{argmax}(M_\mathcal{V}(x_{val})) \right]$. 
$\mathcal{A}$ can evaluate the behavior of $M_{\mathcal{A}}$ compared with $M_{\mathcal{V}}$. 

The two evaluation metrics mentioned above represent different goals of an adversary $\mathcal{A}$.
Accuracy is useful when $\mathcal{A}$ puts importance on performance on unseen test data, e.g., to develop a better service with the substitute model $M_{\mathcal{A}}$ by exploiting the victim model $M_{\mathcal{V}}$. 
On the other hand, fidelity is useful when $\mathcal{A}$ wants to mimic the behavior of $M_{\mathcal{V}}$. 

\subsection{Difficulty of Data-free Model Extraction on Tabular Data} \label{difficultyofdatafree}

Our finding attack, TEMPEST, is a model extraction attack against tabular data models under a reasonable data-free setting. 
As described the detail of TEMPEST in the next section, we describe the technical problems of the previous attacks under data-free settings ~\cite{kariyappa2020maze,truong2021datafree} below. 

The primary problem of the previous attacks is that data normalization is not considered. 
In those settings, it is assumed that an adversary can automatically normalize query data in the range of 0--1, which is identical to a victim model. 
However, in general, the adversary may not know how data is normalized, and thus the assumption in the previous attacks significantly narrows the scope of attacks applicable in the real world. 
For instance, In the case of images processing, features are in the range of 0--255 as pixel values or are naturally given by formats such as JPEG and PNG.
In contrast, tabular data deals with various features, e.g., body height and blood pressure. 
It indicates that, in tabular data models, a range of data may differ for each feature. 


If the adversary normalizes the data without knowledge about their ranges, the performance of the substitute model will no longer be improved unless the normalization process matches that of the victim model. 
The problems mentioned above make the previous data-free attacks challenging to apply to systems in the real world. 

In this paper, to solve these problems, we examine the hypothesis that the performance of a data-free model extraction attack can be improved by using publicly available statistical information. 
Roughly speaking, an adversary can execute a model extraction attack under data-free settings when she can access statistical information such as mean or a pair of maximum and minimum input values that the victim model may use in training.
For example, in some countries, one can utilize statistical information such as mean annual income as financial data or mean body height and weight for each age group and gender as healthcare data. 
We presume that the use of such statistical information can be practical and realistic for significant varieties of tabular data. 

\section{TEMPEST Attack} 

In this section, we present a model extraction attack called TEMPEST (Tabular Extraction from Model by Public and External STatistics).
As mentioned above, by leveraging publicly available statistical information, TEMPEST can realize a substitute model with performance comparable with PRADA~\cite{PRADA}, which is an existing attack with training samples, only by querying to the victim model, i.e., under the data-free setting. 
We first describe an overview of TEMPEST and then explain its algorithms in detail. 

\begin{figure}[t]
    \centering
    \includegraphics[width=0.47\textwidth]{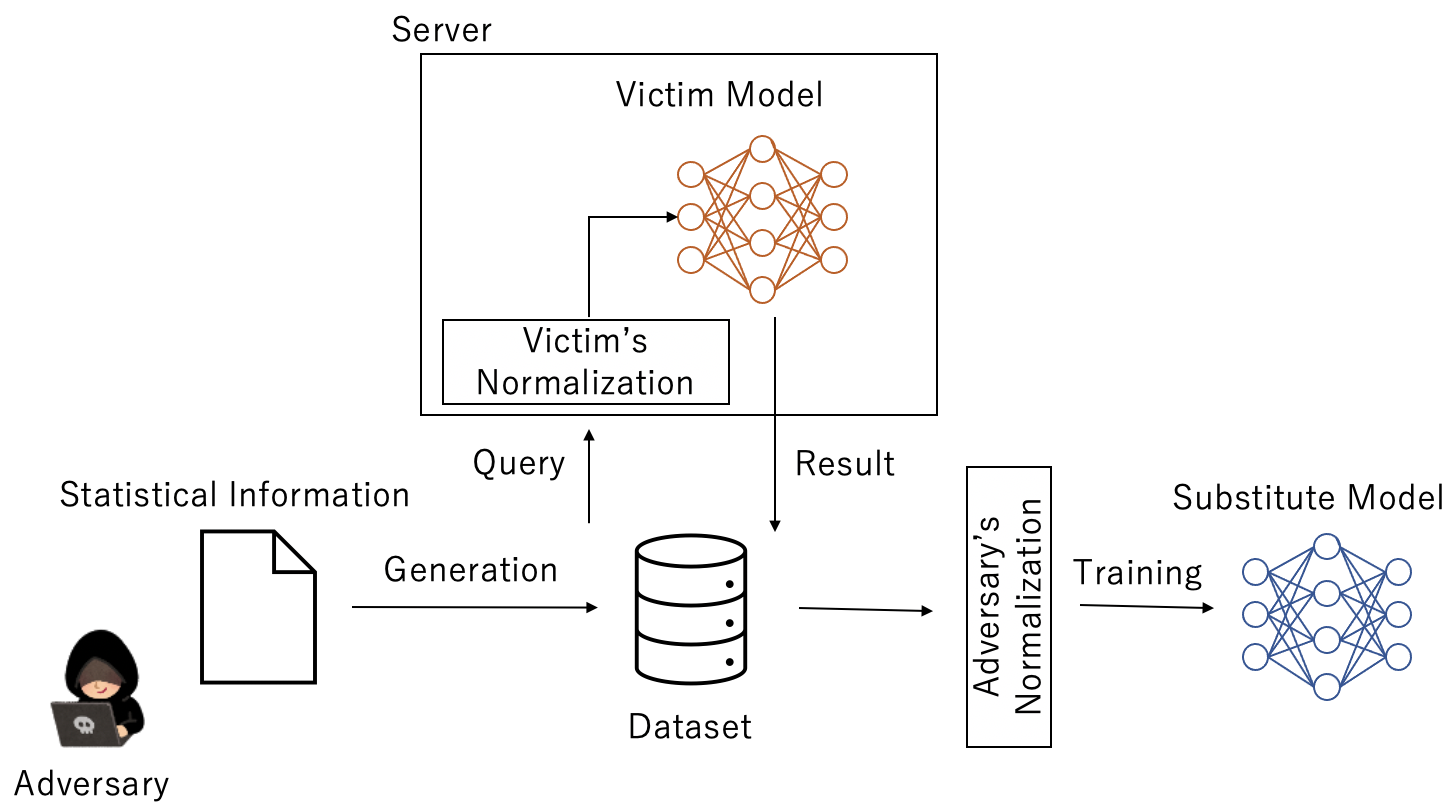}
    \caption{Overview of TEMPEST Attack}
    \label{fig:TEMPEST_overview}
\end{figure}

\subsection{Overview}

Figure \ref{fig:TEMPEST_overview} shows an overview of TEMPEST.
The main idea is the generation of query data to a victim model via statistical information before the training of a substitute model. 

First, in TEMPEST, an adversary generates query data to obtain input-output pairs for training her substitute model. 
Specifically, for each feature values presumably used in training of the victim model, she generates query data from publicly available statistical information (referred to as public information for short).
For instance, the statistical information includes the mean, the variance, or the pair of the minimum and the maximum of the feature values. 
Consequently, the adversary can prepare a suitable dataset for querying the victim model. 
Next, the adversary queries the victim model with the dataset and then trains the substitute model with the queries and their inference results. 

Through the process mentioned above, the adversary can execute the model extraction via APIs of the victim model. 
In doing so, the adversary need no knowledge about the normalization process of data for the victim model. 
Regarding the normalization for the substitute model, the adversary can choose the most effective normalization process by trying multiple normalization processes simultaneously under the public information obtained by the adversary. 
Consequently, the adversary can optimize the training of the substitute model independently of the victim model's normalization process and can execute a model extraction against tabular data models under a realistic setting.

\begin{algorithm}[t]
  \caption{Data Generation of TEMPEST}
  \label{alg:TEMPEST1}
  \begin{algorithmic}[1]
    \Require Victim model $M_{\mathcal{V}}$, substitute model $M_{\mathcal{A}}$, statistical information $X_{\mathcal{P}}$, data generation $gen\_mode$. 
    \Ensure Adversary's training data $X_{\mathcal{A}}$.
    
    
    \State $stat\_params \leftarrow \Call{Select\_public\_knowledge}{X_\mathcal{P}}$
    \If{$gen\_mode ==$ "Gen\_var"}
        \State $mean,var \leftarrow stat\_params$
        \State $X_\mathcal{A} \leftarrow Gaussian(mean,var)$
    \ElsIf{$gen\_mode ==$ "Gen\_min"}
        \State $min,max \leftarrow stat\_params$
        \State $X_\mathcal{A} \leftarrow Uniform(min,max)$
    \EndIf
    \State \textbf{return} $X_\mathcal{A}$
  \end{algorithmic}
\end{algorithm} 
\begin{algorithm}[t]
  \caption{Training of Substitute Model}
  \label{alg:TEMPEST2}
  \begin{algorithmic}[1]
    \Require Victim model $M_{\mathcal{V}}$, substitute model $M_{\mathcal{A}}$, statistical information $X_{\mathcal{P}}$,  adversary's training data  $X_{\mathcal{A}}$, normalization parameter $S_\mathcal{V}$ of the victim model.
    \Ensure Substitute model $M_{\mathcal{A}}$. 
    \Procedure{Query}{$M_{\mathcal{V}}, X_{\mathcal{A}}$}
    \State $X^*_{\mathcal{A}} \leftarrow$ $Victim\_normalization$($X_{\mathcal{A}}, S_\mathcal{V}$)
    \State \textbf{return} $M_{\mathcal{V}}(X^*_{\mathcal{A}})$
    \EndProcedure
    \State $c_{\mathcal{A}} \leftarrow \Call{Query}{M_{\mathcal{V}},X_{\mathcal{A}}}$
    \State $X^{**}_{\mathcal{A}} \leftarrow Adversary\_normalization(X_{\mathcal{A}})$ 
    \State $D_{\mathcal{A}} \leftarrow (X^{**}_{\mathcal{A}}, c_{\mathcal{A}})$
    \State $M_{\mathcal{A}} \leftarrow$ $Training$($M_{\mathcal{A}},D_{\mathcal{A}}$)
    \State \Return $M_{\mathcal{A}}$
  \end{algorithmic}
\end{algorithm}

\subsection{Attack Method}

Here, we describe the details of TEMPEST's attack method. 
TEMPEST is composed of two parts: data generation and training of the substitute model.

Algorithm~\ref{alg:TEMPEST1} shows the data generation process.
In this algorithm, the adversary generates the training data for the substitute model by leveraging the public information.
There are two ways to do this: using the mean and the variance data (denoted by Gen\_var), or using the minimum and the maximum values (denoted by Gen\_min).
The Gen\_var process corresponds to lines 2--4.
In the Gen\_var process, the adversary obtains the mean $mean$ and variance $var$ for each feature from the public information and then generates data under their normal distribution.
On the other hand, the Gen\_min process corresponds to lines 5--7.
In the Gen\_min process, the adversary obtains the minimum $min$ and maximum $max$ for each feature from the public information and then generates data under the uniform distribution in the range of $min$ and $max$. 
The generated data described above are utilized for the training in Algorithm~\ref{alg:TEMPEST2}. 

Next, Algorithm~\ref{alg:TEMPEST2} shows the training algorithm for the substitute model.
The adversary queries the victim model with the generated data in lines 1--4.
Note that, although the adversary does not know the victim model's normalization process $Victim\_scale$ from the assumption, lines 1--4 of the algorithm are executed by the victim model. 
Meanwhile, on line 7, the adversary determines the substitute model's normalization process $Adversary\_normalization$, i.e., the use of the mean and the variance values (denoted by Standard) or the use of the minimum and the maximum values (denoted by MinMax). 
Then, on line 8, the adversary generates a training dataset for the substitute model by utilizing inference results from the victim model. 
The adversary utilizes this dataset to train a substitute model.


\begin{table}[t]
  \caption{Datasets and accuracy of baseline models }
  \label{tab:dataset}
  \centering
  \scalebox{0.8}{
  \begin{tabular}{lrrrr}
    \hline
	Dataset & Features & Instances & Classes & Accuracy \\ \hline \hline 
	Adult & 14 & 32600 & 2 & 81.7\%\\ \hline
	Cancer & 32 & 569 & 2 & 98.9\%\\ \hline
	Diabetes & 9 & 768 & 2 & 76.1\%\\ \hline
	Arrhythmia & 279 & 452 & 16 & 64.3\%\\ \hline
  \end{tabular}
  }
\end{table}

\section{Experiments}  
\label{experiments}

In this section, we evaluate TEMPEST through extensive experiments and then discuss the properties of TEMPEST based on experimental results.
The primary purpose of the experiments is to identify the effectiveness of TEMPEST: more precisely, we confirm whether an adversary executing TEMPEST can obtain a substitute model with high accuracy than existing attacks~\cite{truong2021datafree,kariyappa2020maze,PRADA}. 
Especially, compared with the previous data-free attack~\cite{truong2021datafree}, we demonstrate that TEMPEST can provide the same-level accuracy and fidelity despite overcoming the problem described in Section of Difficulty of Data-free Model Extraction on Tabular Data.


\subsection{Experimental Settings}
\label{experimental setting}

\subsubsection{Datasets and Model Architectures}  

Table~\ref{tab:dataset} shows the datasets used in the experiments. 
Adult\footnote{\url{https://archive.ics.uci.edu/ml/datasets/adult}}, Cancer\footnote{\url{https://archive.ics.uci.edu/ml/datasets/Breast+Cancer+Wisconsin+(Diagnostic)}}, and Diabetes\footnote{\url{https://www.kaggle.com/uciml/pima-indians-diabetes-database}} 
are tabular datasets for two-class classifications.
As a more complex task, we use Arrhythmia\footnote{\url{https://archive.ics.uci.edu/ml/datasets/arrhythmia}} as a dataset with a sixteen-class classification. 
For numerical variables, query data to a victim model are generated via statistical information. 
The Adult dataset contains several features as categorical variables (also known as a nominal variable), and a datum is randomly chosen from instances for each of such categorical features. 
Each dataset is divided into 1:3:1 into the training of the victim model, validation of the substitute model, and publicly available statistical information for the data generation of TEMPEST. 

We utilize a fully connected three-layer neural network with 90 nodes in the hidden layer for all datasets as a model architecture. 
Hyper-parameters for each model are as follows: learning rates for the victim and substitute models are 0.01, and the number of epochs is 30. 
The mean and variance are utilized for the data normalization process on the victim and substitute models.


\subsubsection{Baseline} 
The accuracies of the victim models for the datasets described above are 81.7\%, 98.9\%, 76.1\%, and 64.3\% for Adult, Cancer, Diabetes, and Arrhythmia, respectively. 
These are used as baselines. 
We also compare the accuracy and fidelity of TEMPEST with PRADA~\cite{PRADA} as a typical model extraction attack. 
PRADA ~\cite{PRADA} assumes that an adversary has several training data of a victim model as initial samples. 
The following experiments assume that an adversary executing PRADA has 10 samples for each class as initial knowledge except for Arrhythmia. 
For Arrhythmia, 76 samples are utilized as initial knowledge because the number of samples for each class on Arrhythmia is unbalanced. 
The normalization process are executed for these samples on the victim model. 
We also compare the accuracy and the fidelity of TEMPEST with those of a data-free attack, whereby query data are randomly generated from 0--1 as the normalized data without generative adversarial networks in the existing works~\cite{kariyappa2020maze,truong2021datafree}. 
Hereafter, we simply call the above method the data-free for the sake of convenience.



We note that the following experiments are based on the divided datasets as mentioned above, and experiments with publicly available statistical information in the real world are presented in Section of Real-World Case Study. 

\subsection{Experimental Results}

\begin{figure*}[t]
    \centering
    \begin{tabular}{cccc}
        \begin{minipage}{0.25\hsize}
            \begin{center}
            \includegraphics[scale=0.2]{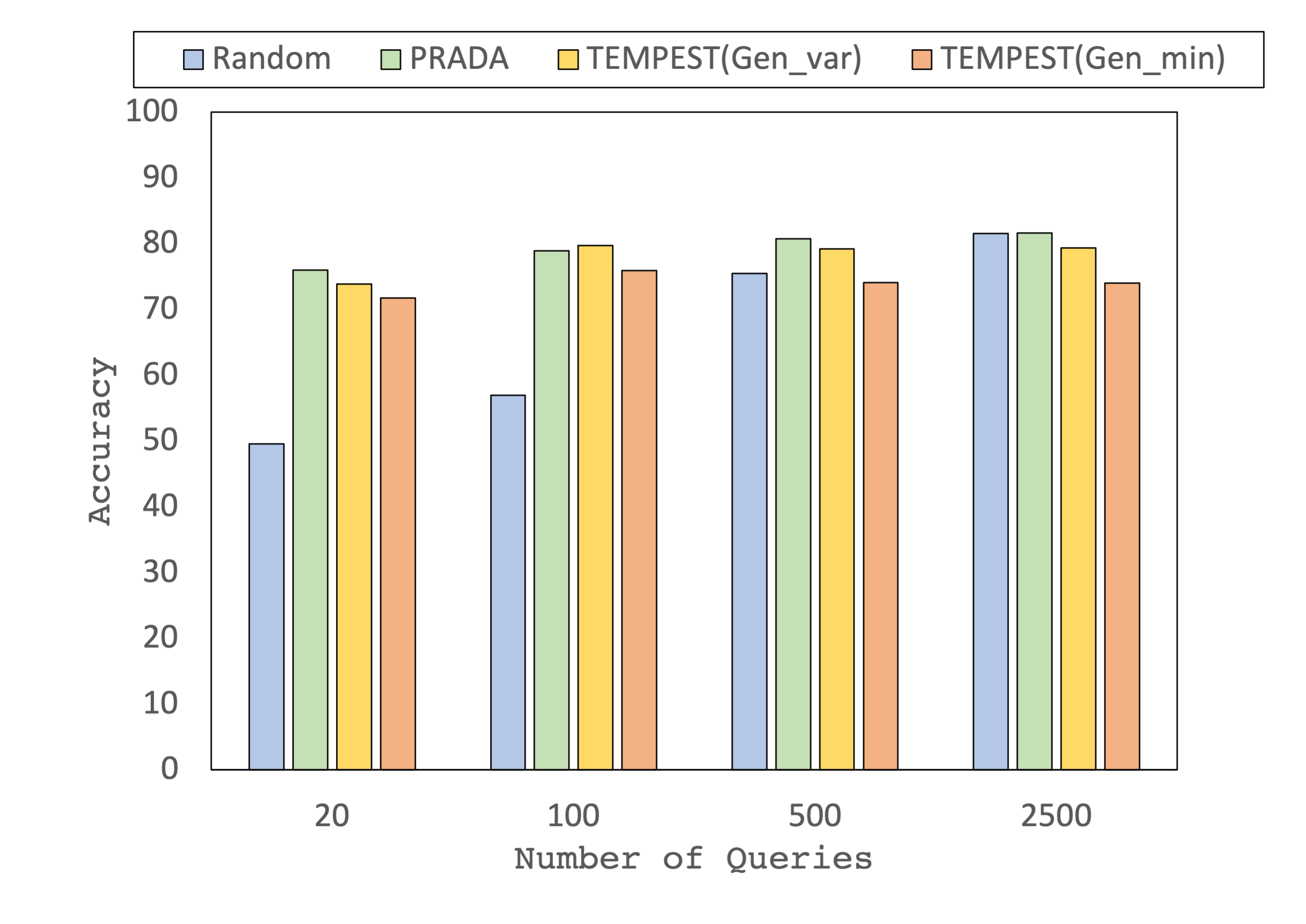}
            \end{center}
            \subcaption{Adult}
        \end{minipage}
        \begin{minipage}{0.25\hsize}
            \begin{center}
            \includegraphics[scale=0.2]{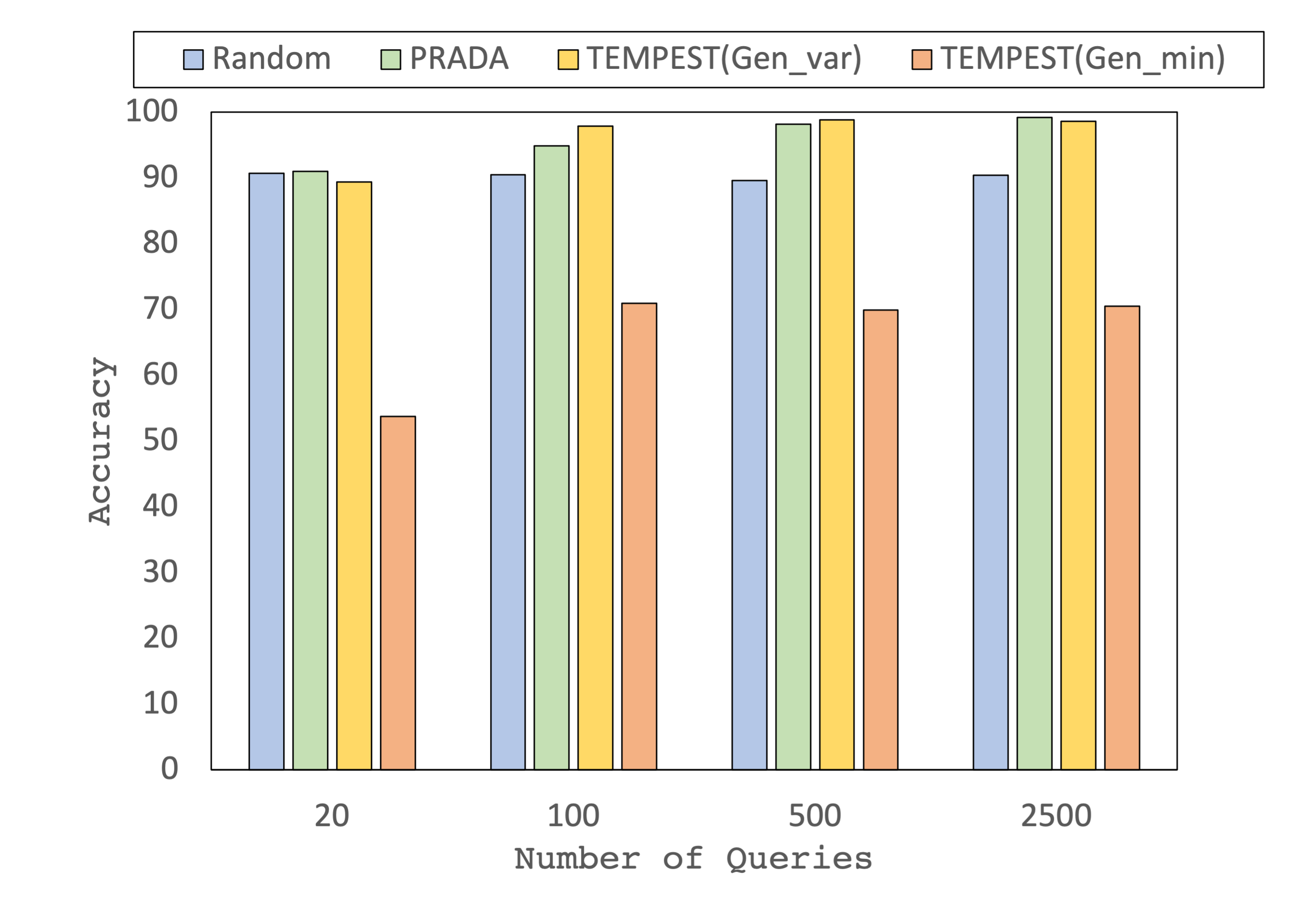}
            \end{center}
            \subcaption{Cancer}
        \end{minipage}
        
        \begin{minipage}{0.25\hsize}
            \begin{center}
            \includegraphics[scale=0.2]{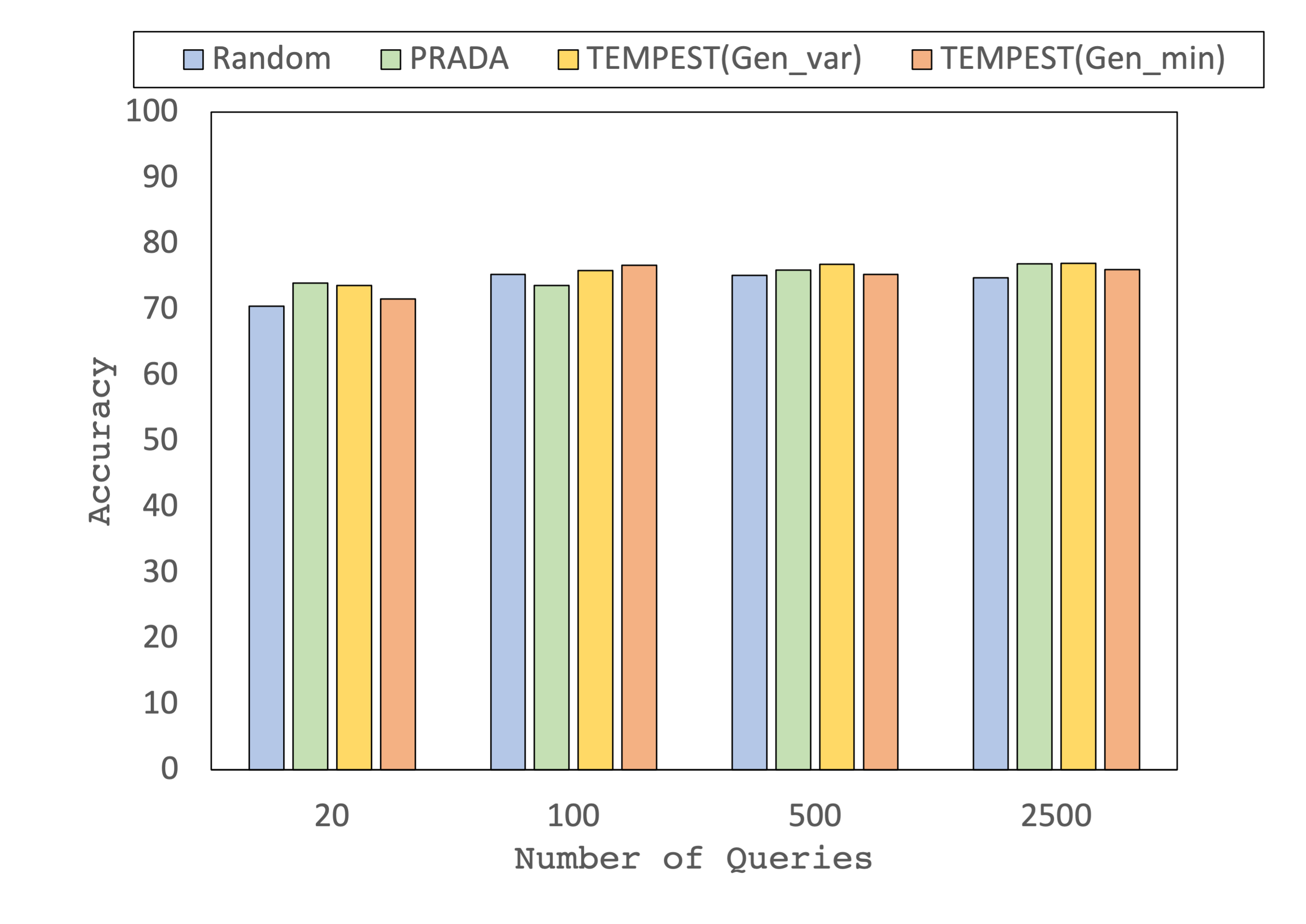}
            \end{center}
            \subcaption{Diabetes}
        \end{minipage}
        \begin{minipage}{0.25\hsize}
            \begin{center}
            \includegraphics[scale=0.2]{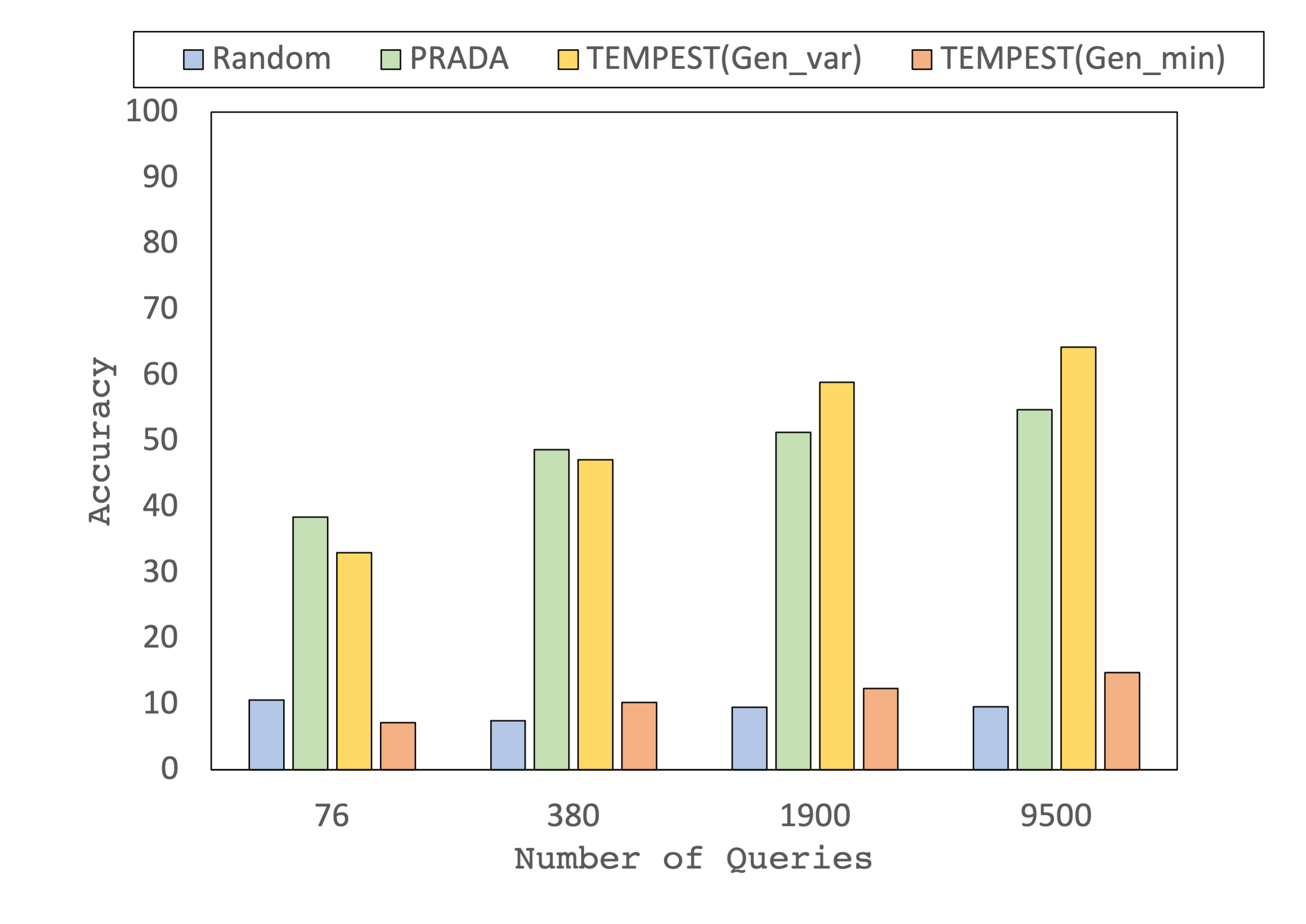}
            \end{center}
            \subcaption{Arrhythmia}
        \end{minipage}
    \end{tabular}
    
    \caption{Accuracy evaluation of substitute models for each dataset}
    \label{fig:acc}
\end{figure*}

\begin{figure*}[t]
    \centering
    \begin{tabular}{cccc}
        \begin{minipage}{0.25\hsize}
        \begin{center}
        \includegraphics[scale=0.2]{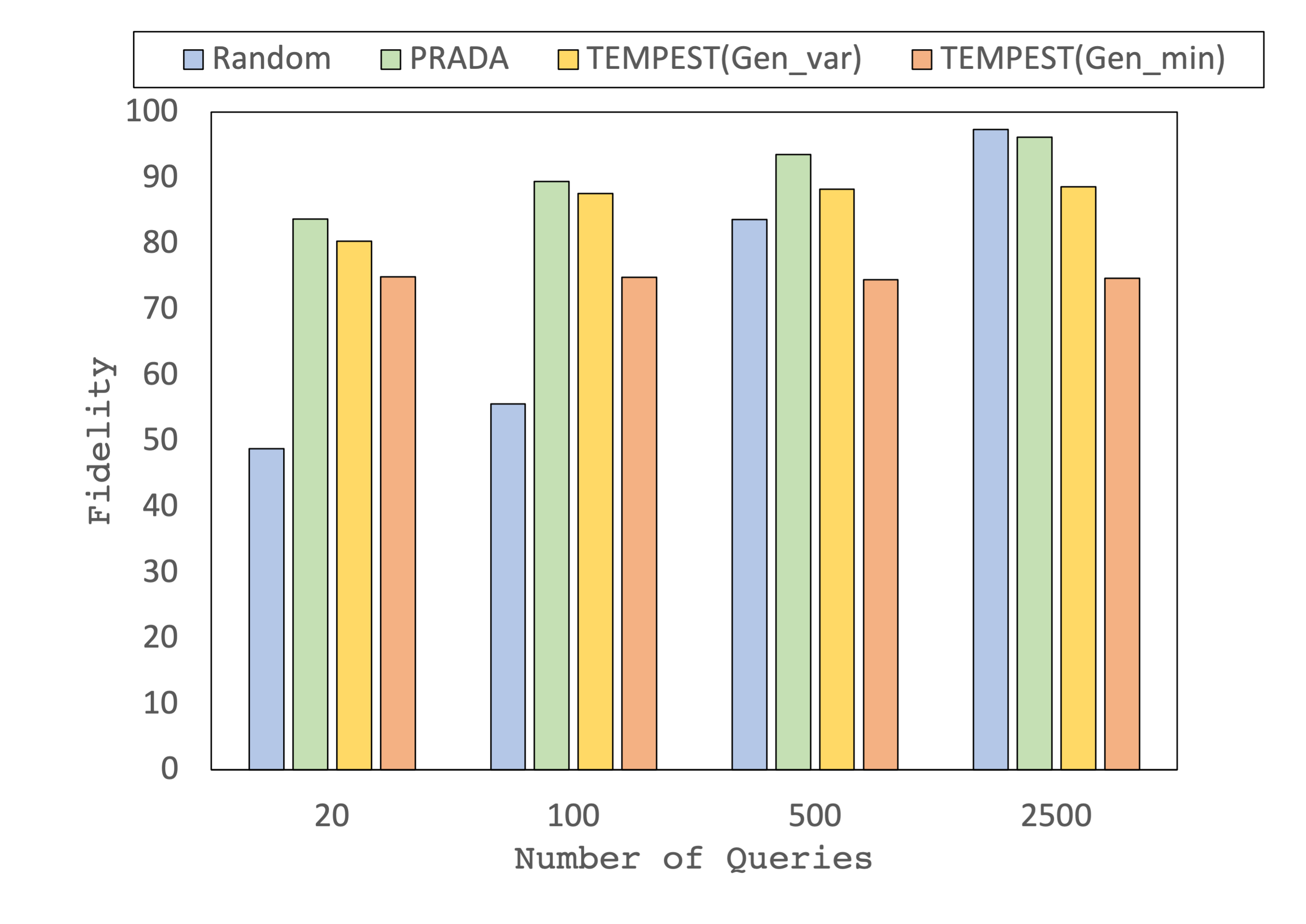}
        \end{center}
        \subcaption{Adult}
    \end{minipage}
    \begin{minipage}{0.25\hsize}
        \begin{center}
        \includegraphics[scale=0.2]{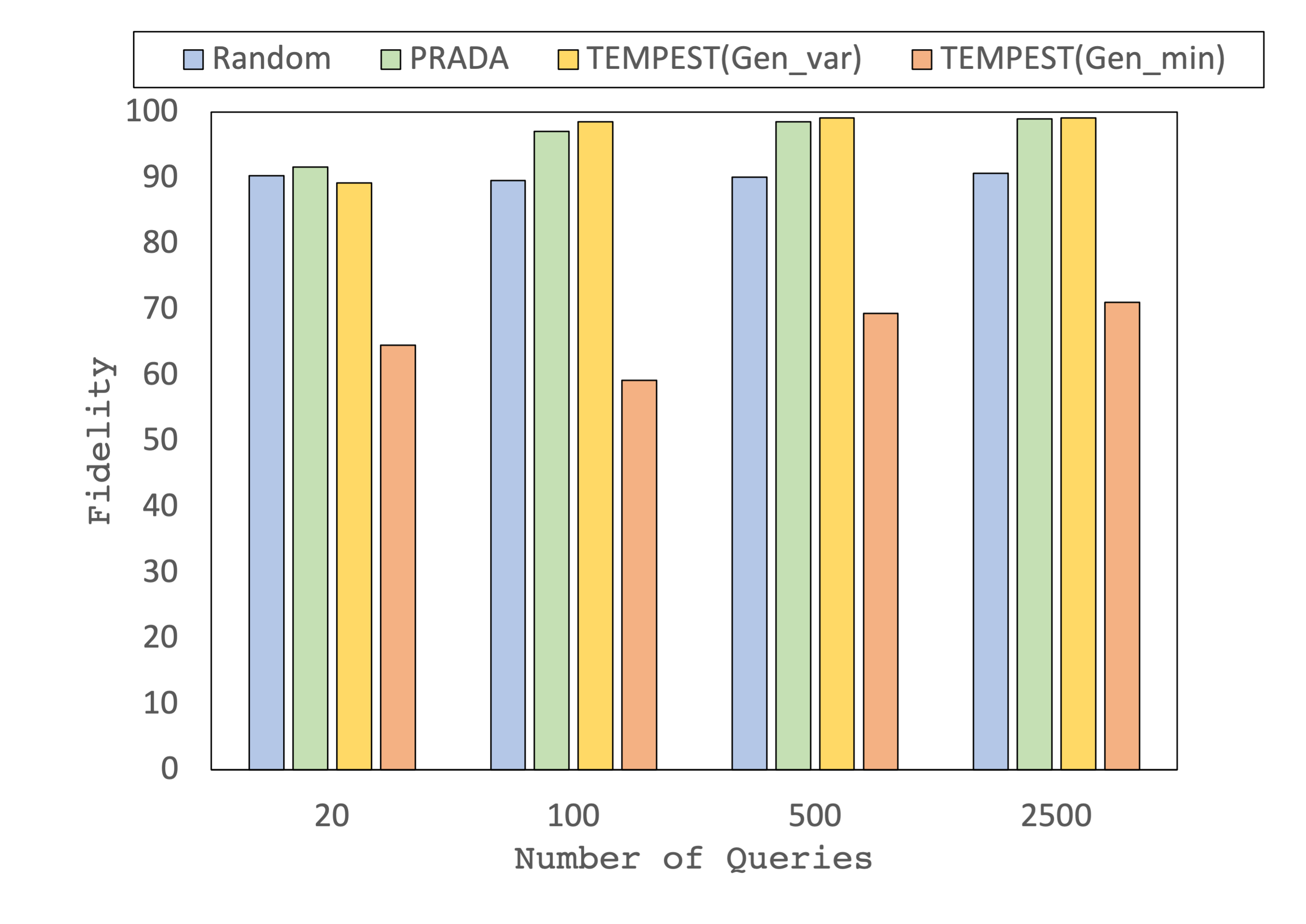}
        \end{center}
        \subcaption{Cancer}
    \end{minipage}
    \begin{minipage}{0.25\hsize}
        \begin{center}
        \includegraphics[scale=0.2]{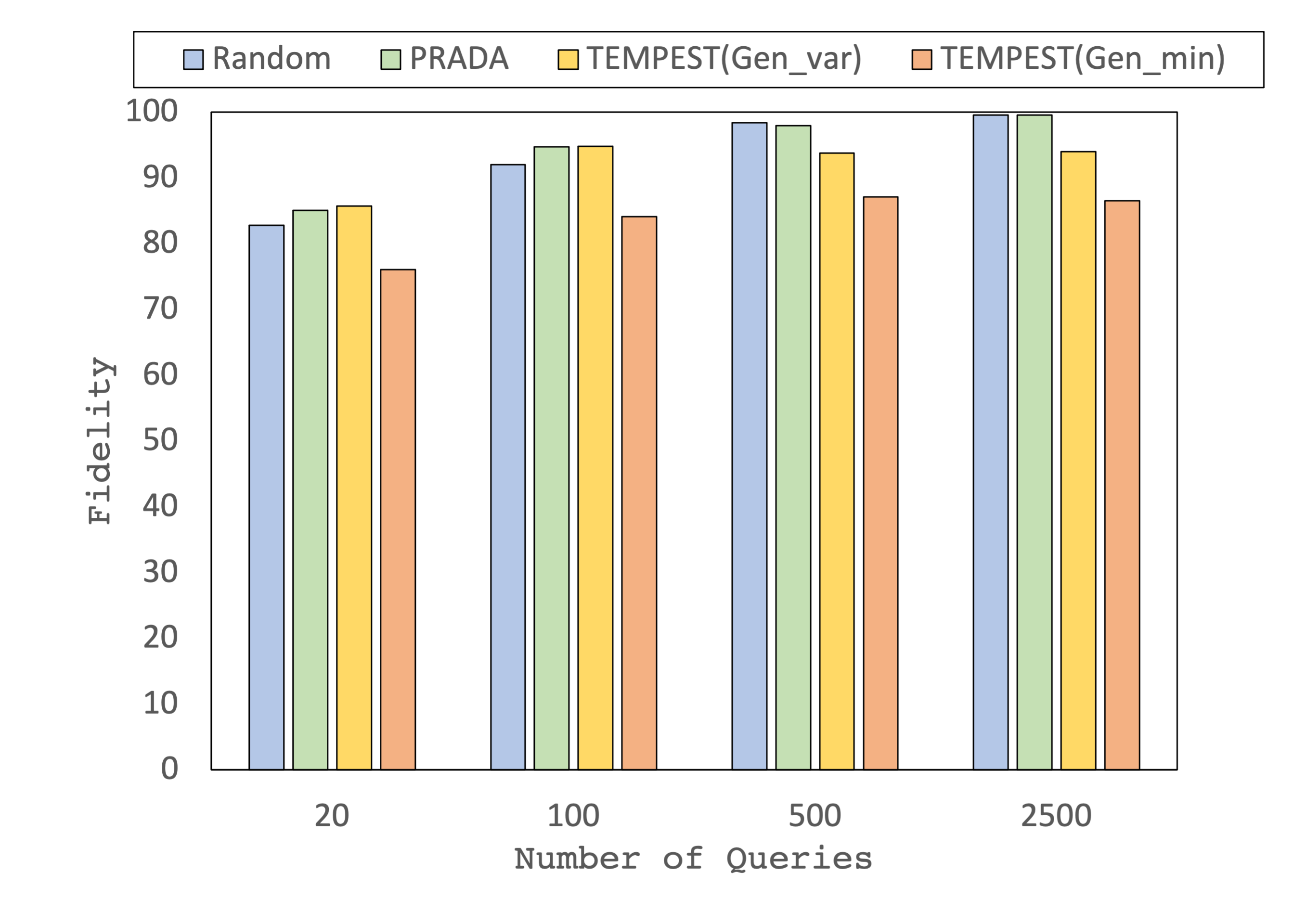}
        \end{center}
        \subcaption{Diabetes}
    \end{minipage}
    \begin{minipage}{0.25\hsize}
        \begin{center}
        \includegraphics[scale=0.2]{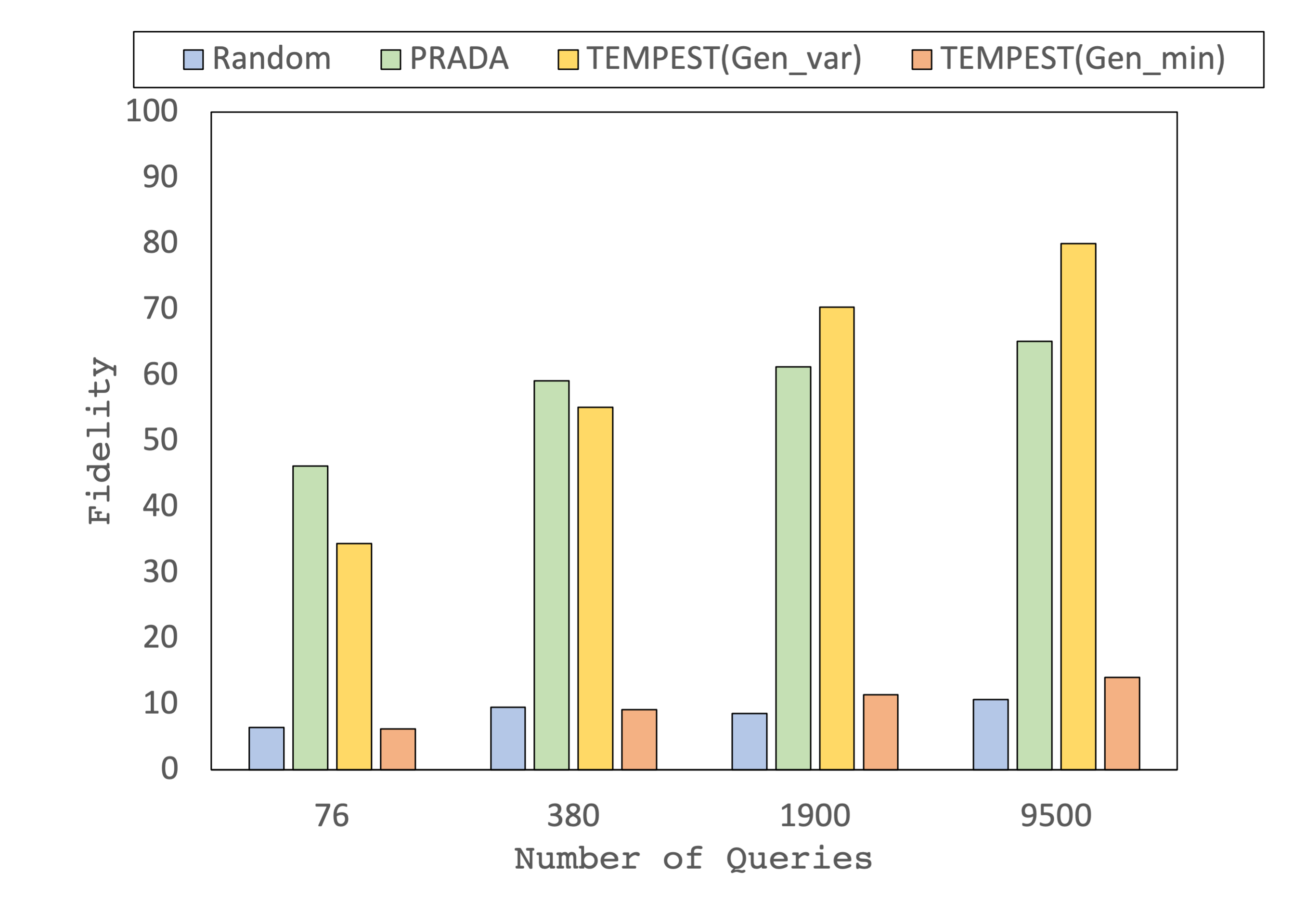}
        \end{center}
        \subcaption{Arrhythmia}
    \end{minipage}
    \end{tabular}

    \caption{Fidelity evaluation of substitute models for each dataset}
    \label{fig:fid}
\end{figure*}

We describe the experimental results below. 
Figure~\ref{fig:acc} and Figure~\ref{fig:fid} show that the accuracy and fidelity of substitute models by PRADA~\cite{PRADA}, the data-free, and TEMPEST, respectively. 



\textbf{Data Generation of TEMPEST:} The accuracy and fidelity of Gen\_var are higher than Gen\_min in all the datasets. We consider that data generated by the mean and the variance covers the distribution of the training data of the victim model than data sampled uniform-randomly. Therefore, it is desirable to use the mean and variance for the data generation as much as possible.

\textbf{Comparison to PRADA:} According to Figure~\ref{fig:acc}, the accuracy of TEMPEST is lower than PRADA when queries correspond to the initial samples. 
Meanwhile, the accuracy difference is within five points, and, consequently, it is considered that the data generation is executed effectively.  
Besides, the accuracies of TEMPEST for Diabetes and Arrhythmia are higher than PRADA in proportion to the number of queries. 
The reason for higher accuracy by PRADA with 2500 queries for Adult and Cancer is that the training of a substitute model by TEMPEST was converged. 

According to Figure~\ref{fig:fid}, 
the fidelity of TEMPEST is lower than PRADA for datasets with a small number of dimensions of features, i.e., Adult and Diabetes. 
By contrast, the fidelity of TEMPEST is higher than PRADA for datasets whose dimensions are large, i.e., Cancer and Arrhythmia. 

Based on the above results, TEMPEST will achieve higher accuracy and fidelity than attacks with initial samples, such as PRADA, on tabular data whose number of instances is limited. 

\textbf{Comparison to Data-Free:} 
For datasets whose dimensions of features are limited, such as Adult and Diabetes, the accuracy of TEMPEST is almost the same as the data-free, even if the number of queries is limited. 
The reason is that the domain of the data generation is limited due to the small dimensions. 
In other words, random generation by the data-free can enough capture the distribution of the training data. 
In contrast, the accuracy and the fidelity by TEMPEST (and PRADA) are significantly higher than the data-free for Arrhythmia, whose dimensions are enormous. 
It is considered that the decision boundary obtained by the data-free is unstable because enough distribution is no longer obtained due to the enormous dimensions. 

\section{Discussion} \label{discussion}

In this section, we discuss the properties and advantages of TEMPEST based on the experimental results in the previous section. 
We first shed light on the effects on the normalization process and the number of dimensions for each dataset. 
Next, we consider the accuracy improvement with a few initial samples and the effect on inference classes. 
We also identify the real-world case study, limitation, and potential countermeasures of TEMPEST. 

\subsection{Normalization Process of Substitute Model}

\begin{figure}[t]
    \centering
    \begin{tabular}{cc}
        \begin{minipage}{0.47\hsize}
            \includegraphics[scale=0.15]{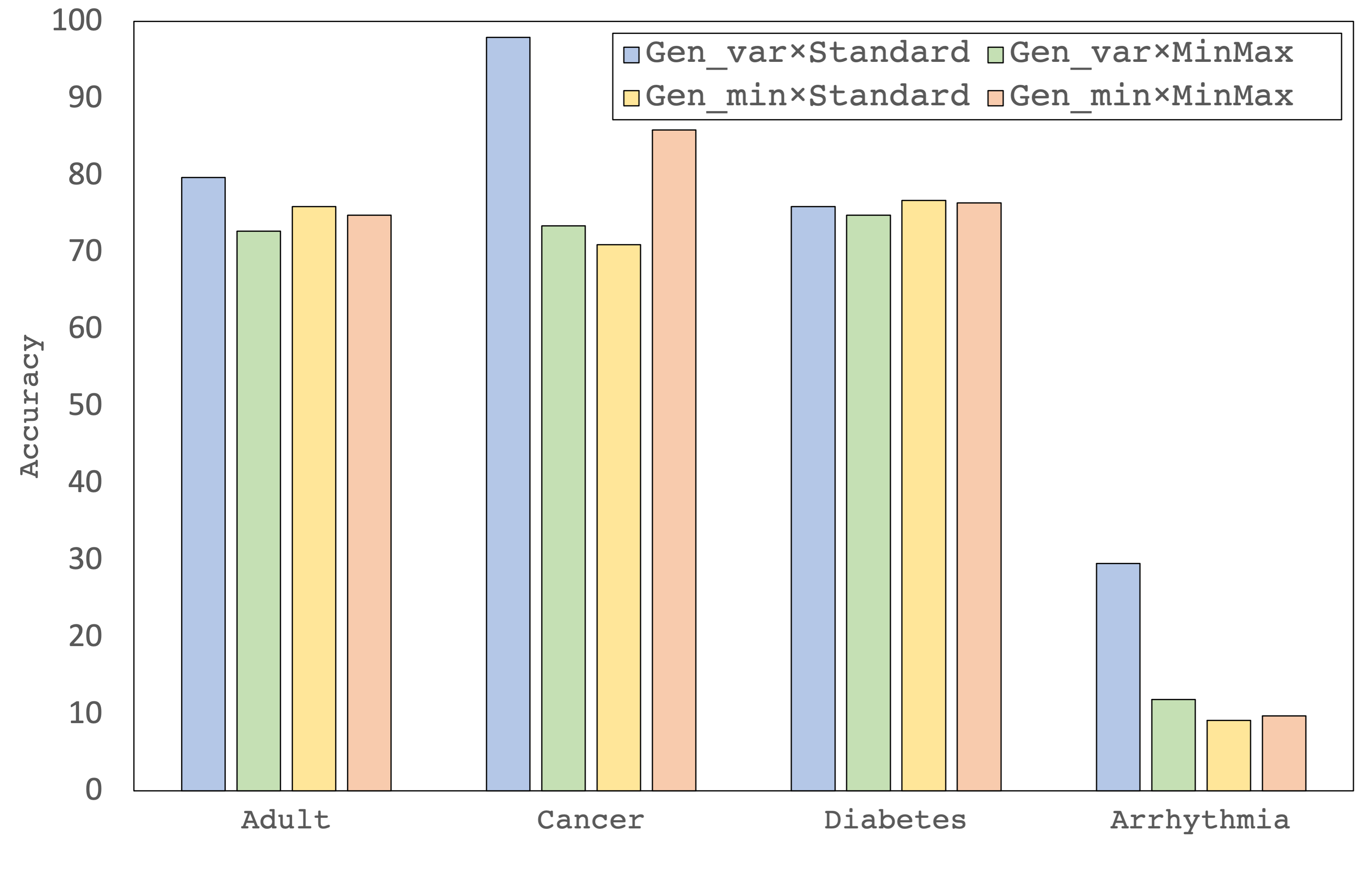}
            \subcaption{Standard}
            \label{fig:standard}
        \end{minipage}
        \hfill
        \begin{minipage}{0.47\hsize}
            \includegraphics[scale=0.15]{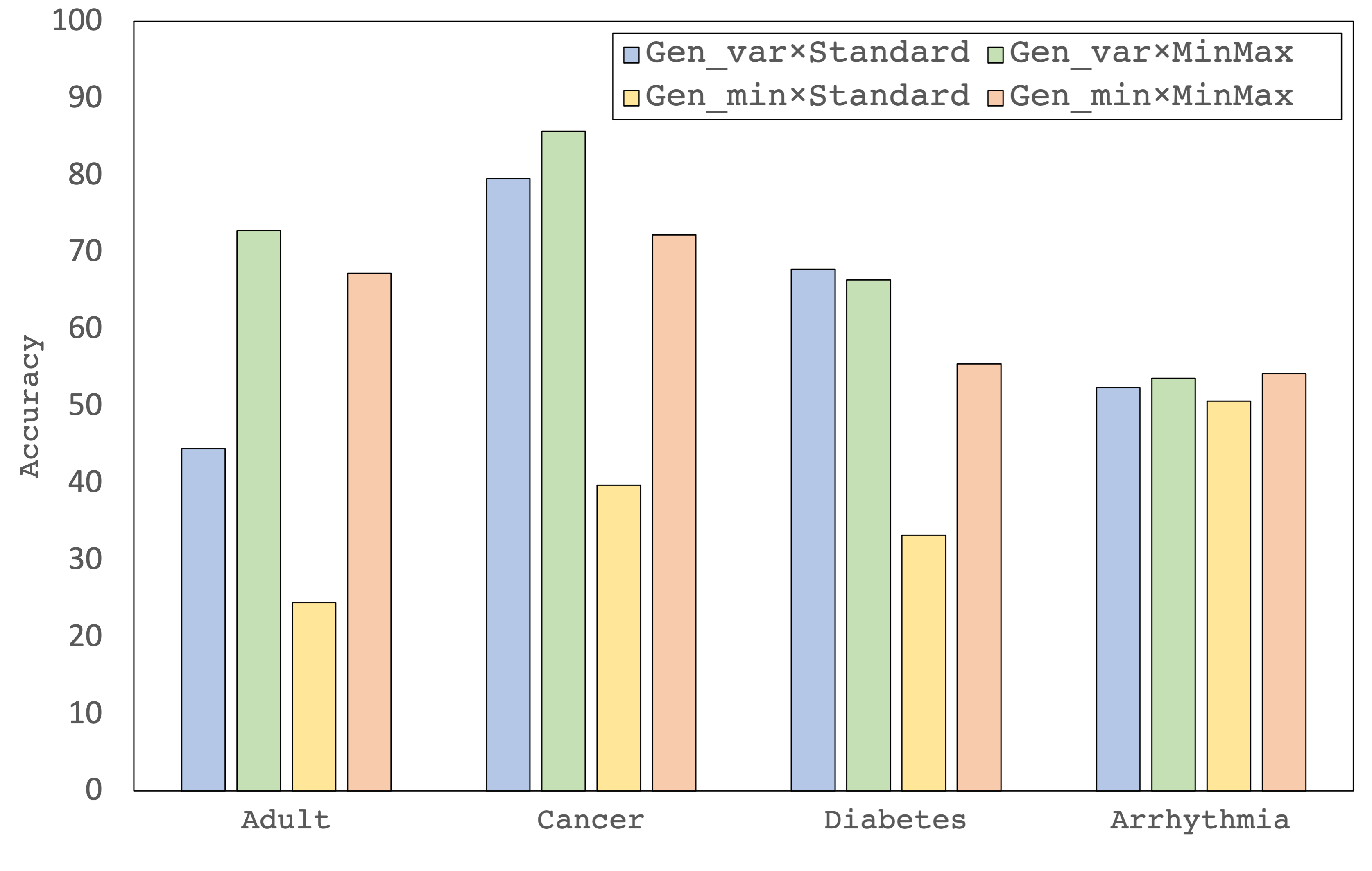}
            \subcaption{MinMax}
            \label{fig:minmax}
        \end{minipage}
    \end{tabular}
    \caption{Accuracy evaluation of substitute models for each normalization process of the victim model}
\end{figure}

We discuss the effect on the normalization process of a substitute model. 
In particular, on a substitute model of TEMPEST, the accuracy for each normalization process is evaluated. 
Then, the result to a victim model with Standard, i.e., the use of the mean and the variance, as the normalization process is shown in Figure~\ref{fig:standard} and that with MinMax as the normalization process is shown in Figure~\ref{fig:minmax}, respectively. 
Here, each graph is measured by utilizing a pair of mean and variance, denoted by Gen\_var for the sake of convenience, or a pair of minimum and maximum values, denoted by Gen\_min for the sake of convenience, as the query data generation of TEMPEST, respectively. 

According to Figure~\ref{fig:standard}, utilizing mean and variance for both the query data generation and the normalization process, i.e., the case of Gen\_var$\times$Standard, provided high accuracies. 
This means that higher accuracy can be obtained stably by using mean and variance for the query data generation and the same normalization process as the victim model. 


\subsection{Effect on Initial Samples}
We identify whether the accuracy and fidelity of a substitute model on TEMPEST is improved if an adversary has initial samples. 
Specifically, suppose that an adversary owns 10\% samples as initial samples in the same setting described in Section of Experimental Settings and the remaining 90\% samples are generated by TEMPEST. 
In doing so, the initial samples are equally divided for each inference class. 
For instance, when the number of samples are twenty, a single sample is given in each class, i.e., totally two samples for two classes. 
Likewise, five samples is given in each class for 1000 queries while 25 samples is given in each class for 5000 queries. 
The number of initial samples on Arrhythmia are eleven samples for 76 queries, 48 samples for 380 queries, and 113 samples for 1900 queries due to the bias of samples for each class. 
Based on the above setting, let TEMPEST without initial samples be a baseline. 


The result with initial samples is presented in Table~\ref{tab:init_acc_fed}. 
Except for Diabetes, the accuracy and fidelity of the substitute model were improved for all the dataset by the initial samples when the number of queries is limited. 
Notably, about five points were improved for both metrics on Cancer and Arrhythmia. 
In contrast, except for Arrhythmia, the difference from the baseline becomes smaller for all the datasets in proportion to the number of queries. 

Based on the above result, the accuracy and fidelity of TEMPEST can be improved by owning initial samples, especially when the number of queries is limited. 
Intuitively, the training with initial samples is effective if the number of queries is small. 
By contrast, the advantage of initial samples is less since the training of a substitute model is converged by increasing the number of queries. 
For data with limited instances, e.g., Arrhythmia, the accuracy and fidelity can be improved in proportion to initial samples. 


\begin{table*}[t]
    \centering
    \caption{Evaluation of accuracy and fidelity of TEMPEST with initial samples}
    \scalebox{0.9}{
    \begin{tabular}{ccccccccc}\hline
         & \multicolumn{2}{c}{Adult} & \multicolumn{2}{c}{Cancer} & \multicolumn{2}{c}{Diabetes} & \multicolumn{2}{c}{Arrhythmia}\\ \hline
        Queries & Accuracy [\%] & Fidelity [\%] & Accuracy [\%] & Fidelity [\%] & accuracy [\%] & Fidelity [\%] & Accuracy [\%] & Fidelity [\%] \\ \hline
        20(76) & 74.43(+0.59) & 81.41(+1.03) & 94.25(+4.91) & 93.74(+4.52) & 72.51(-1.13) & 86.42(+0.72) & 38.23(+5.25) & 40.47(+6.06)\\ \hline
        100(380) & 78.90(-0.82) & 88.33(+0.72) & 98.21(+0.33) & 98.21(-0.32) & 76.53(+0.63) & 94.03(-0.76) & 52.74(+5.59) & 65.00(+9.88) \\ \hline
        500(1900) & 79.16(-0.03) & 88.30(+0.02) & 98.84(+0.01) & 99.15(+0.06) & 76.69(-0.13) & 94.09(+0.31) & 60.58(+1.15) & 77.03(+6.67)\\ \hline
    \end{tabular}
    }
    \label{tab:init_acc_fed}
\end{table*}



\subsection{Effect on Number of Inference Classes} \label{multi-class}

We evaluate the effect on the number of inference classes for a victim model. 
Since most tabular datasets are two classes, MNIST is evaluated as a typical multi-class classification, although image classification. 
Similar to Section of Experimental Settings, the accuracy and fidelity of a substitute model are evaluated by comparing with PRADA~\cite{PRADA}. The results are shown in Table~\ref{tab:mnist_acc_fid}. 

\begin{table*}[t]
  \caption{Accuracy and fidelity of a substitute model on the MNIST evaluation}
  \label{tab:mnist_acc_fid}
  \centering
  \scalebox{0.85}{
  \begin{tabular}{ccccccc}
    \hline
	 & \multicolumn{2}{c}{\shortstack{PRADA~\cite{PRADA} \\ with initial samples }} & \multicolumn{2}{c}{\shortstack{PRADA~\cite{PRADA} \\ without initial samples}} & \multicolumn{2}{c}{TEMPEST} \\ \hline 
	Queries  & Accuracy [\%] & Fidelity [\%] & Accuracy [\%] & Fidelity [\%] & Accuracy [\%] & Fidelity [\%] \\ \hline
	300 & 33.49  & 33.91 & 10.32 & 10.42 & 10.02 & 9.95 \\ \hline
	1500 & 41.44 & 72.80 & 50.39 & 13.23 & 10.06 & 9.95 \\ \hline
	7500 & 78.48 & 89.82 & 52.79 & 17.86 & 16.62 & 16.76 \\ \hline
    37500 & 82.38 & 95.83 & 54.77 & 24.88 & 35.40 & 36.20 \\ \hline
\end{tabular}
}
\end{table*}

\begin{figure}[t]
    \centering
    \includegraphics[scale=0.2]{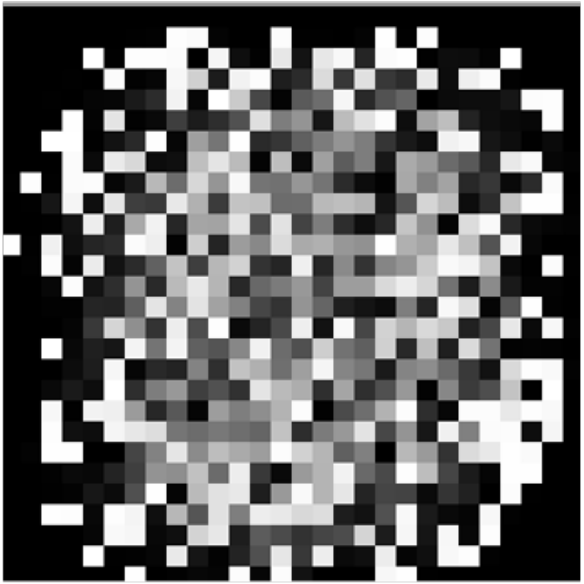}
    \caption{Figure generated by TEMPEST}
    \label{fig:mnist}
\end{figure}

According to Table~\ref{tab:mnist_acc_fid}, the accuracy and fidelity of TEMPEST are lower than PRADA on the MNIST evaluation. 
Such a negative factor is caused because statistical information is often ineffective for images, and thus the metrics become lower. 
For instance, an example generated by TEMPEST is shown in Figure~\ref{fig:mnist}. 
The example generated by the statistical information is obliviously far from any digit. 

Nevertheless, it is also considered that multi-class for MNIST decreases the accuracy and fidelity of a substitute model. 
Multi-class inference requires an adversary to gather more statistical information for each class. 
Indeed, the fidelity on Arrhythmia with 16 classes is lower than the other datasets with two classes as shown in Figure~\ref{fig:fid}. 
Further studies, which take improving TEMPEST for multi-class inference into account, will need to be undertaken. 


\subsection{Real-World Case Study} \label{real-world}

We discuss the effectiveness of TEMPEST in a real-world scenario below. 
We specifically identify whether model extraction is successful against a victim model with a dataset described in the previous section by leveraging publicly available statistical information in the real world. 

\begin{figure}[t]
    \centering
    \begin{tabular}{cc}
        \begin{minipage}{0.47\hsize}
            \includegraphics[scale=0.2]{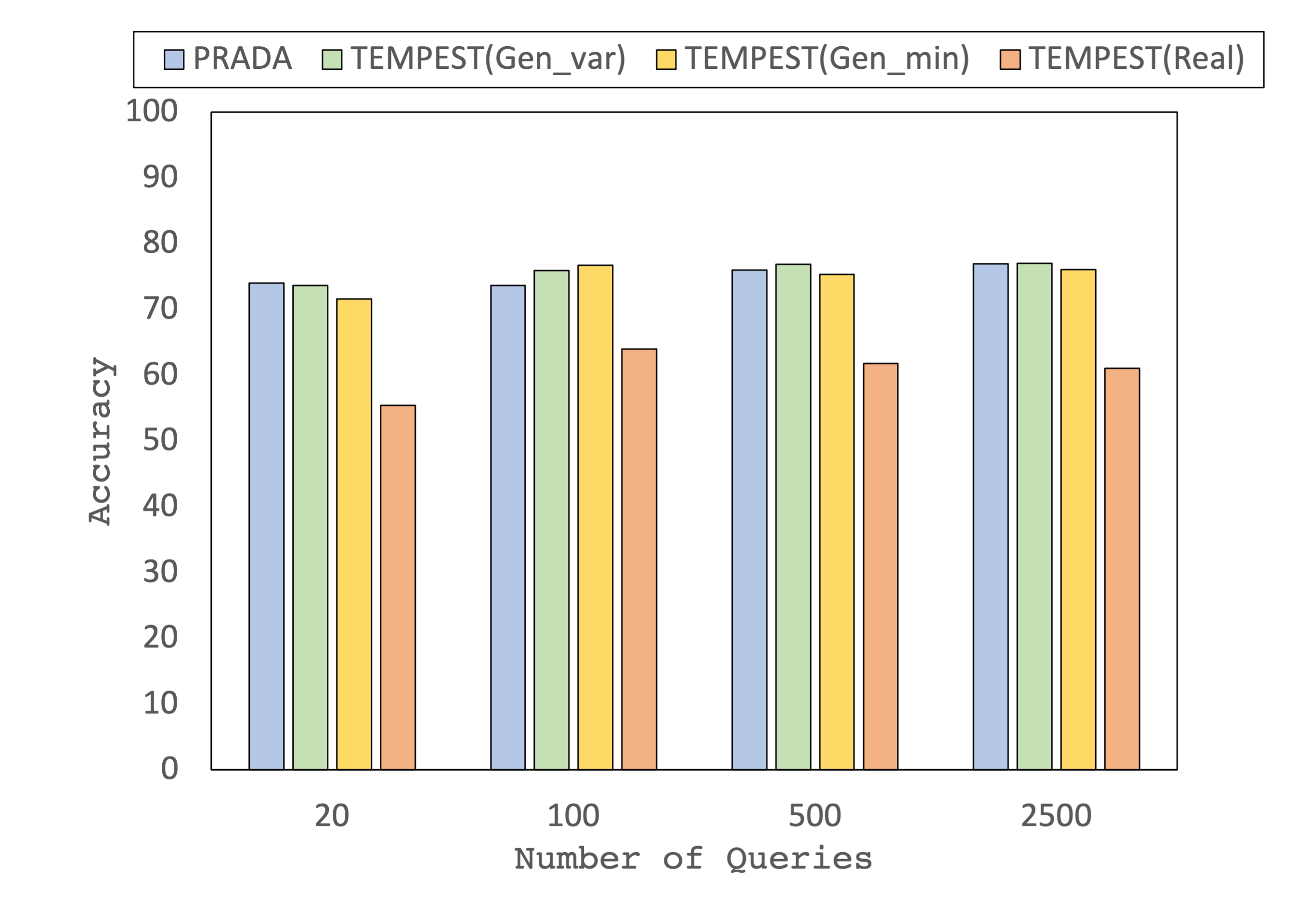}
            \subcaption{Accuracy}
            \label{fig:realdiabetesacc}
        \end{minipage}
        \begin{minipage}{0.47\hsize}
            \includegraphics[scale=0.2]{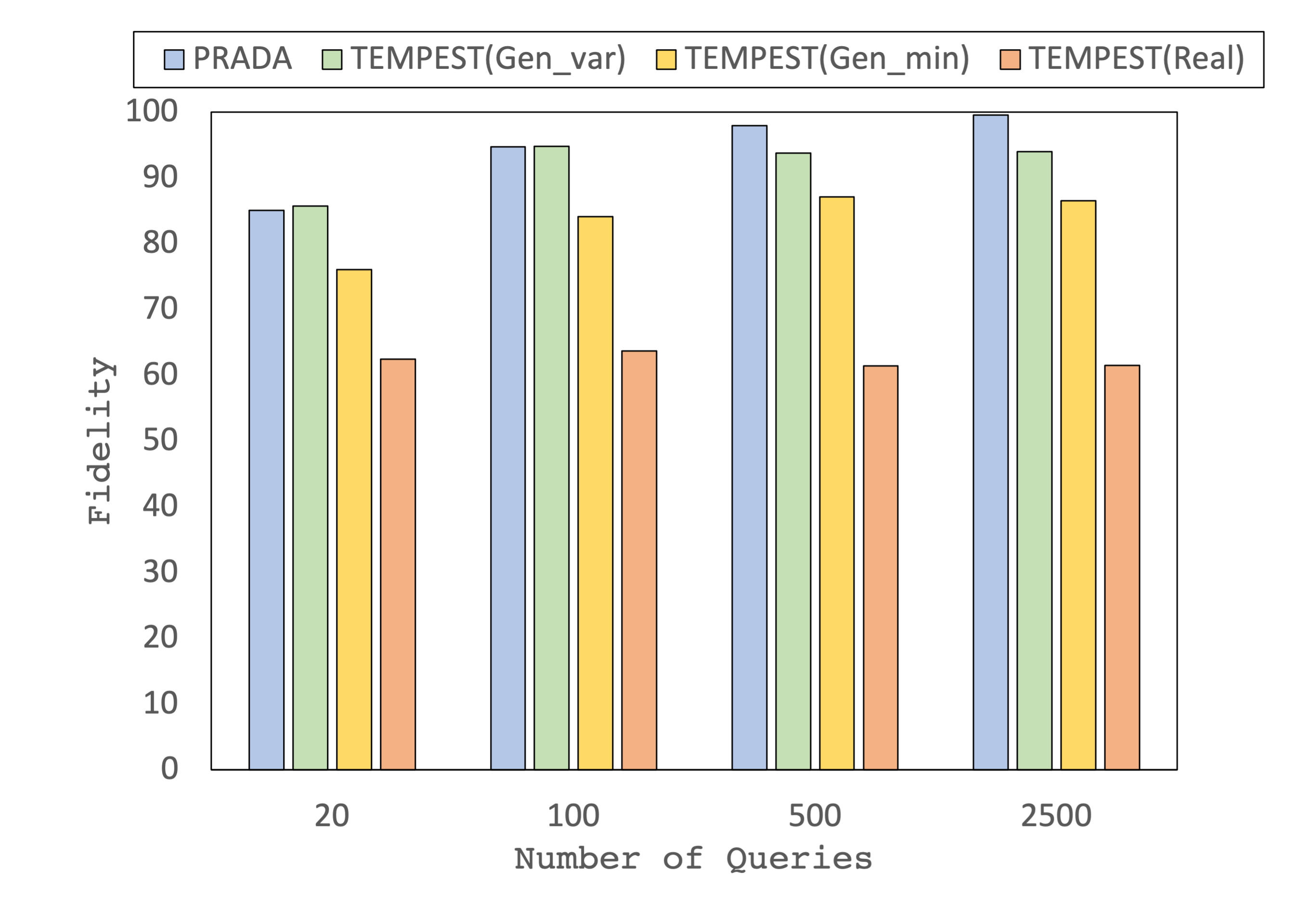}
            \subcaption{Fidelity}
            \label{fig:realdiabetesfid}
        \end{minipage}
    \end{tabular}
    \caption{Real-world case study for Diabetes}
    \label{fig:Diabetesinrealworld}
\end{figure}

As a survey on existing public data, for eight features on the Diabetes dataset, we found mean and variance of five features, i.e., glucose, blood pressure, insulin, BMI, age, for diabetes-positive patients from existing medical data~\cite{Kim2008PimaIndian,Tataranni1998PimaIndian}. 
Query data are then generated by randomly sampling in the range of 0--1 for the remaining three features. 
The result in the setting described above is shown in Figure~\ref{fig:Diabetesinrealworld}. 
On the figure, whereas TEMPEST (Real) corresponds to the above result, PRADA, TEMPEST (Gen\_var), and TEMPEST (Gen\_min) correspond to the setting in Section of Experiments. 

According to the figure, the accuracy and fidelity on TEMPEST (Real) deteriorate fifteen points and thirty points compared to those on TEMPEST (Gen\_var) and PRADA. 
The foremost cause of these deteriorations is the use of mean and variance for the positive patients. 
As analyzing the detail of results on the substitute model, queries related to positive patients are primarily generated from the substitute model, and inference results for the positive patients are returned from the victim model. 
In doing so, normalization parameters $stat\_params$ are different for the substitute model and the victim model. 
Data for negative patients on the victim model are then normalized periodically to the range of positive patients on the substitute model. 
Consequently, the substitute model mistakenly infers with respect to the negative patients, and then the accuracy and fidelity deteriorated. 

Although the above performance was not ideal, we nevertheless believe that TEMPEST can provide a substitute model with enough accuracy and fidelity against a victim model whose inference classes have a bias for each inference class, as long as mean and variance for the class are available. 
Indeed, the substitute model could infer all the positive patients precisely on the results mentioned above of Diabetes. 
It means that the recall for positive patients is 100\%. 
Also, TEMPEST would provide high accuracy and fidelity if mean and variance for negative patients are available. 

The above results also indicate that TEMPEST is potentially effective even for a general model that has no bias for each inference class, as long as its mean and variance are publicly available. 
Further tests will confirm our insights mentioned above of TEMPEST. 

\subsection{Limitation and Potential Countermeasures}

We discuss limitation and potential countermeasures for TEMPEST below. 
The most important limitation of TEMPEST is that an adversary need publicly available statistics. 
It indicates that, for example, executing the attack is difficult against a victim model whose training data is unpublished, e.g., sensitive data for users. 
Likewise, the attack would be brutal if statisticsal information are meaningless, as shown in the results of MNIST. 

Moreover, as shown in Section of Effect on Number of Inference Classes, the amount of the statistical information for an adversary should be large if dimensions of features or target classes are major. 
We consider that the accuracies deteriorated due to a lack of statistical information because MNIST and Arrhythmia include many classes. 

A potential countermeasure against TEMPEST is differential privacy~\cite{Dwork06} for the training data. 
Whereas BDPL~\cite{zheng2019bdpl} which applies differential privacy to inference by a victim model has been proposed as an existing countermeasure, we consider applying it to the training as a potential countermeasure against TEMPEST. 
In general, differential privacy disturbs the training data distribution and, for example, mean and variance become different from the original distribution. 
As a result, the training data of a victim model has a gap from the public statistical information, and thus the data generation in Algorithm~\ref{alg:TEMPEST1} will no longer reach a workable level. 
To further our research, we plan to investigate the detail of the above countermeasure. 

\section{Conclusion} 

In this paper, we presented a novel model extraction attack, TEMPEST, by considering data normalization on tabular data. 
By assuming the use of public statistical information, TEMPEST enables an adversary to execute a model extraction attack without training data, i.e., in the data-free setting, under a realistic setting. 
Based on the experimental results with tabular datasets for dimensions of features and the number of instances, we demonstrated that mean and variance are effective as statistical information for the data generation of TEMPEST. 
Moreover, we identified that the normalization process of a substitute model should be identical to that of a victim model. 
Furthermore, in a case that the number of instances is limited, the same-level of accuracy and fidelity as the existing precise attack~\cite{PRADA} with initial samples could be achieved in the data-free setting by virtue of TEMPEST. 
We plan to release the source code via GitHub for reproducibility and a reference to subsequent works. 
We believe that TEMPEST can realize a substitute model with high accuracy and fidelity under data-free setting in the real world. 
Future work will concentrate on a countermeasure against TEMPEST by utilizing differential privacy without the accuracy deterioration. 


\bibliographystyle{aaai22.bst}
\bibliography{main} 

\begin{thebibliography}{26}
\providecommand{\natexlab}[1]{#1}

\bibitem[{Arik and Pfister(2021)}]{arik2021tabnet}
Arik, S.~O.; and Pfister, T. 2021.
\newblock TabNet: Attentive Interpretable Tabular Learning.
\newblock In \emph{Proc. of AAAI 2021}, volume~35, 6679--6687. AAAI.

\bibitem[{Chandrasekaran et~al.(2020)Chandrasekaran, Chaudhuri, Giacomelli,
  Jha, and Yan}]{chandrasekaran2020exploring}
Chandrasekaran, V.; Chaudhuri, K.; Giacomelli, I.; Jha, S.; and Yan, S. 2020.
\newblock Exploring Connections Between Active Learning and Model Extraction.
\newblock In \emph{Proc. of {USENIX} Security 2020}, 1309--1326. {USENIX}
  Association.

\bibitem[{Chen et~al.(2021)Chen, Guo, Zhang, Xie, and
  Liu}]{chen2021stealingrein}
Chen, K.; Guo, S.; Zhang, T.; Xie, X.; and Liu, Y. 2021.
\newblock Stealing Deep Reinforcement Learning Models for Fun and Profit.
\newblock In \emph{Proc. of ASIACCS 2021}, 307–319. ACM.
\newblock ISBN 9781450382878.

\bibitem[{Dwork(2006)}]{Dwork06}
Dwork, C. 2006.
\newblock Differential Privacy.
\newblock In \emph{Proc. of ICALP 2006}, volume 4052 of \emph{LNCS}, 1--12.
  Springer.

\bibitem[{Hu et~al.(2020)Hu, Liang, Li, Deng, Zuo, Ji, Xie, Ding, Liu,
  Sherwood, and Xie}]{hu2020deepsniffer}
Hu, X.; Liang, L.; Li, S.; Deng, L.; Zuo, P.; Ji, Y.; Xie, X.; Ding, Y.; Liu,
  C.; Sherwood, T.; and Xie, Y. 2020.
\newblock DeepSniffer: A DNN Model Extraction Framework Based on Learning
  Architectural Hints.
\newblock In \emph{Proc. of ASPLOS 2020}, 385–399. ACM.

\bibitem[{Hua, Zhang, and Suh(2018)}]{hua2018reverse}
Hua, W.; Zhang, Z.; and Suh, G.~E. 2018.
\newblock Reverse Engineering Convolutional Neural Networks Through
  Side-channel Information Leaks.
\newblock In \emph{Proc. of DAC 2018}, 1--6. IEEE.

\bibitem[{Juuti et~al.(2019)Juuti, Szyller, Marchal, and Asokan}]{PRADA}
Juuti, M.; Szyller, S.; Marchal, S.; and Asokan, N. 2019.
\newblock PRADA: Protecting against DNN Model Stealing Attacks.
\newblock In \emph{Proc. of EuroS\&P 2019}, 512--527. IEEE.

\bibitem[{Kariyappa, Prakash, and Qureshi(2021)}]{kariyappa2020maze}
Kariyappa, S.; Prakash, A.; and Qureshi, M.~K. 2021.
\newblock MAZE: Data-Free Model Stealing Attack Using Zeroth-Order Gradient
  Estimation.
\newblock In \emph{Proc. of CVPR 2021}, 13814--13823.

\bibitem[{Keskar et~al.(2020)Keskar, McCann, Xiong, and
  Socher}]{shirish2020thievesmonolingual}
Keskar, N.~S.; McCann, B.; Xiong, C.; and Socher, R. 2020.
\newblock The Thieves on Sesame Street are Polyglots - Extracting Multilingual
  Models from Monolingual APIs.
\newblock In \emph{Proc. of EMNLP 2020}, 6203--6207. ACL.

\bibitem[{Kim et~al.(2008)Kim, Pavkov, Looker, Nelson, Bennett, Hanson, Curtis,
  Sievers, and Knowler}]{Kim2008PimaIndian}
Kim, N.~H.; Pavkov, M.~E.; Looker, H.~C.; Nelson, R.~G.; Bennett, P.~H.;
  Hanson, R.~L.; Curtis, J.~M.; Sievers, M.~L.; and Knowler, W.~C. 2008.
\newblock Plasma Glucose Regulation and Mortality in Pima Indians.
\newblock \emph{Diabetes Care}, 31(3): 488--492.

\bibitem[{Krishna et~al.(2019)Krishna, Tomar, Parikh, Papernot, and
  Iyyer}]{Krishna2020Thieves}
Krishna, K.; Tomar, G.~S.; Parikh, A.~P.; Papernot, N.; and Iyyer, M. 2019.
\newblock Thieves on Sesame Street! Model Extraction of BERT-based APIs.
\newblock \emph{CoRR}, abs/1910.12366: 1--18.

\bibitem[{Orekondy, Schiele, and Fritz(2019)}]{orekondy2019knockoff}
Orekondy, T.; Schiele, B.; and Fritz, M. 2019.
\newblock Knockoff Nets: Stealing Functionality of Black-Box Models.
\newblock In \emph{Proc. of CVPR 2019}, 4954--4963. IEEE.

\bibitem[{Pal et~al.(2020)Pal, Gupta, Shukla, Kanade, Shevade, and
  Ganapathy}]{pal2020activethief}
Pal, S.; Gupta, Y.; Shukla, A.; Kanade, A.; Shevade, S.~K.; and Ganapathy, V.
  2020.
\newblock ActiveThief: Model Extraction Using Active Learning and Unannotated
  Public Data.
\newblock In \emph{Proc. of AAAI 2020}, volume~34, 865--872. AAAI.

\bibitem[{Papernot et~al.(2017)Papernot, McDaniel, Goodfellow, Jha,
  Berkay~Celik, and Swami}]{Papernot2017}
Papernot, N.; McDaniel, P.; Goodfellow, I.; Jha, S.; Berkay~Celik, Z.; and
  Swami, A. 2017.
\newblock Practical Black-Box Attacks Against Machine Learning.
\newblock In \emph{Proc. of ASIACCS 2017}, 506--519. ACM.

\bibitem[{Reith, Schneider, and Tkachenko(2019)}]{reith2019efficiently}
Reith, R.~N.; Schneider, T.; and Tkachenko, O. 2019.
\newblock Efficiently Stealing Your Machine Learning Models.
\newblock In \emph{Proc. of WPES 2019}, 198–210. ACM.

\bibitem[{Szegedy et~al.(2014)Szegedy, Zaremba, Sutskever, Bruna, Erhan,
  Goodfellow, and Fergus}]{szegedy2014intriguing}
Szegedy, C.; Zaremba, W.; Sutskever, I.; Bruna, J.; Erhan, D.; Goodfellow,
  I.~J.; and Fergus, R. 2014.
\newblock Intriguing properties of neural networks.
\newblock In \emph{Proc. of ICLR 2014}.

\bibitem[{Szyller et~al.(2021)Szyller, Duddu, Gr{\"{o}}ndahl, and
  Asokan}]{szyller2021goodartists}
Szyller, S.; Duddu, V.; Gr{\"{o}}ndahl, T.; and Asokan, N. 2021.
\newblock Good Artists Copy, Great Artists Steal: Model Extraction Attacks
  Against Image Translation Generative Adversarial Networks.
\newblock \emph{CoRR}, abs/2104.12623.

\bibitem[{Tataranni et~al.(1998)Tataranni, Christin, Snitker, Paolisso, and
  Ravussin}]{Tataranni1998PimaIndian}
Tataranni, P.~A.; Christin, L.; Snitker, S.; Paolisso, G.; and Ravussin, E.
  1998.
\newblock {Pima Indian Males Have Lower $\beta$-Adrenergic Sensitivity Than
  Caucasian Males}.
\newblock \emph{The Journal of Clinical Endocrinology \& Metabolism}, 83(4):
  1260--1263.

\bibitem[{Tram{\'e}r, Zhang, and Juels(2016)}]{FFA16}
Tram{\'e}r, F.; Zhang, F.; and Juels, A. 2016.
\newblock Stealing Machine Learning Models via Prediction APIs.
\newblock In \emph{Proc. of USENIX Security 2016}, 601--618. USENIX
  Association.

\bibitem[{Truong et~al.(2021)Truong, Maini, Walls, and
  Papernot}]{truong2021datafree}
Truong, J.-B.; Maini, P.; Walls, R.~J.; and Papernot, N. 2021.
\newblock Data-Free Model Extraction.
\newblock In \emph{Proc. of CVPR 2021}, 4771--4780.

\bibitem[{Wang and Zhenqiang~Gong(2018)}]{wang2018stealinghyper}
Wang, B.; and Zhenqiang~Gong, N. 2018.
\newblock Stealing Hyperparameters in Machine Learning.
\newblock In \emph{Proc. of IEEE S\&P 2018}, 36--52. IEEE.

\bibitem[{Yang et~al.(2019)Yang, Dai, Yang, Carbonell, Salakhutdinov, and
  Le}]{yang2019xlnet}
Yang, Z.; Dai, Z.; Yang, Y.; Carbonell, J.; Salakhutdinov, R.~R.; and Le, Q.~V.
  2019.
\newblock XLNet: Generalized Autoregressive Pretraining for Language
  Understanding.
\newblock In \emph{Proc. of NeurIPS 2019}, 5753--5763. Curran Associates, Inc.

\bibitem[{Yoon et~al.(2020)Yoon, Zhang, Jordon, and van~der
  Schaar}]{yoon2020vime}
Yoon, J.; Zhang, Y.; Jordon, J.; and van~der Schaar, M. 2020.
\newblock VIME: Extending the Success of Self- and Semi-supervised Learning to
  Tabular Domain.
\newblock In \emph{Proc. of NeurIPS 2020}, volume~33, 11033--11043. Curran
  Associates, Inc.

\bibitem[{Zanella-Beguelin et~al.(2021)Zanella-Beguelin, Tople, Paverd, and
  K{\"o}pf}]{zanella-beguelin21a2021greybox}
Zanella-Beguelin, S.; Tople, S.; Paverd, A.; and K{\"o}pf, B. 2021.
\newblock Grey-box Extraction of Natural Language Models.
\newblock In \emph{Proc. of ICML 2021}, volume 139 of \emph{PMLR},
  12278--12286. PMLR.

\bibitem[{Zheng et~al.(2019)Zheng, Ye, Hu, Fang, and Shi}]{zheng2019bdpl}
Zheng, H.; Ye, Q.; Hu, H.; Fang, C.; and Shi, J. 2019.
\newblock BDPL: A Boundary Differentially Private Layer Against Machine
  Learning Model Extraction Attacks.
\newblock In \emph{Proc. of ESORICS 2019}, volume 11735 of \emph{LNCS}, 66--83.
  Springer.

\bibitem[{Zhu et~al.(2021)Zhu, Cheng, Zhou, and Lu}]{zhu2021hermes}
Zhu, Y.; Cheng, Y.; Zhou, H.; and Lu, Y. 2021.
\newblock Hermes Attack: Steal {DNN} Models with Lossless Inference Accuracy.
\newblock In \emph{Proc. of {USENIX} Security 2021}. {USENIX} Association.

\end{thebibliography}

\end{document}